# A New Embedded-Atom Method Approach Based On the *p*-th Moment Approximation

Kun Wang[a,b,d], Wenjun Zhu[b*], Shifang Xiao[c], Jun Chen[d], Wangyu Hu[a*]

[a] *College of Materials Science and Engineering, Hunan University, Changsha 410082, P. R. China*

[b] *Laboratory for Shock Wave and Detonation Physics, Institute of Fluid Physics, Mianyang 621900, P. R. China*

[c] *Department of Applied Physics, Hunan University, Changsha 410082, P. R. China*

[d] *Laboratory of Computational Physics, Institute of Applied Physics and Computational Mathematics, Beijing 100088, PR China*

## Abstract

Large scale atomistic simulations with suitable interatomic potentials are widely employed by scientists or engineers of different areas. Quick generation of high-quality interatomic potentials is of urgent need under present circumstances, which largely relies on the developments of potential construction methods and algorithms in this area. Quantities of interatomic potential models have been proposed and parameterized with various methods, such as analytic method, force-matching approach and multi-object optimization method, in order to make the potentials more transferable. Without apparently lowing precisions for describing the target system, potentials of fewer fitting parameters (FPs) are somewhat more physically reasonable. Thus, studying methods of reducing FP number is helpful to understand the underline physics of simulated systems and generalize the construction methods to other similar systems. However, few reported works concentrate on methods of reducing the number of FPs without affecting precisions. In this work, the methods of reducing the FP number while keeping the precisions are discussed from two aspects. Firstly, the physical ideas of constructions of the embedded-atom method (EAM) potential model are modified to make the potential more robust, flexible and scalable without introducing too many FPs. The new EAM potential consists of a new manybody term, based on the *p*-th moment approximation to the tight binding theory and the general transformation invariance of EAM potentials, and an energy modification term represented by pairwise interactions. The pairwise interactions are evaluated by an analytic-numerical scheme without the need to know their functional forms *a priori*. Validations of the new approach are demonstrated via constructing three potentials of aluminum and comparing with a commonly used EAM potential model. Our results show that the new EAM potential needs fewer FPs and smaller cutoff distance to simulate mechanical behaviors of aluminum than some reported potentials. Secondly, the smaller reference data set (SRD), compared with the target reference database for describing the mechanical behaviors of aluminum, is employed in our construction procedures. Through studying sensitivities of extrapolated quantities to uncertainties of the SRD, we find that the extrapolated results could match with the reference data by slightly adjusting the values of low-precision quantities in the SRD rather than adding additional FPs. Additionally, a commonly used EAM model is found to be comparable with the new model at the vicinities of equilibrium states, whose empirical parameter in its embedding term could be related to the effective order of moments of local density of states. This relationship facilitates us to analysis the precisions of this type of EAM models when using them to develop potentials for other metals.

I. INTRODUCTION

With the promoting computation capabilities of modern computers and its broadening applications in scientific and engineering fields, ultra-/large atomic simulations have been proven to be an effective tool for researches of a wide range of scientific and engineering fields due to its high calculation efficiencies and reasonable precisions, such as nano-thermodynamics [1], biology [2], geophysics [3], material physics [4-6] and compression science [7-13]. Upon applying to solve a certain problem, the atomistic simulations should be combined with proper interatomic potentials. Rapid developments of matter research fields require an effective way to construct accurate and transferable interatomic potentials with high efficiencies. Embedded-atom method (EAM) potential proposed by Daw and Baskes [14, 15] is one of the most widely used semi-empirical interatomic potentials, which is based on density functional theory. While another important EAM potential proposed by Finnis and Sinclair [16] is derived from the second moment approximation to tight binding, which is usually named FS potential. Despite of the different theoretical backgrounds, they share a similar form of the energy expression. Different parameterization schemes have been designed to construct the EAM potentials for target applications. Because of complex relationships between EAM potential parameters and resultant behavior of the potential, traditional construction procedure involves a two-step iterative process [17, 18]. First, EAM parameters are determined based on a handful of critical material properties of element in a selected reference structure. Second, the new potential is applied to circumstances not included in the construction phase in order to test its validation and transferability. If the validation is not satisfactory, one need repeat the two steps from the beginning whose construction processes are very tedious and time-consuming. To overcome this drawback, Ercolessi and Adams [19] developed a force-matching method that fit a cubic-spline based potential to ab initio atomic forces of various atomic configurations at finite temperature. Recently, Kim and his coworkers [17] have extended the force-matching method to develop a multi-objective optimization (MOO) procedure for the construction of modified embedded-atom method (MEAM) potentials. The MOO procedure, represented by weighted sum method, could optimally reproduce multiple target values that consist of material properties, and enable one to construct MEAM potentials with minimal manual fittings. The weighted sum method is also employed by other researchers for the construction of different EAM potentials [20-23]. In principle, the method allows one to incorporate arbitrary number of physical properties, whose collection is referred to as reference database set, in the construction procedure of a certain EAM potential. To make EAM functions flexible enough to access to its real ones, more potential parameters are often required for the bigger reference database set which is usually achieved by introducing cubic spline functions instead of the ones based on certain theoretical analyses [19-22]. This method has been proved to be a very powerful approach to EAM potentials of various elements under complex application circumstances, for example simulating properties of the iron–phosphorus system [20], phase transformations in zirconium [24] and the mechanical behaviors of magnesium [17, 25]. However, with the growing complexities of application circumstances (characterized by a large target reference database), the fitting parameter (FP) number of the cubic-spline style potentials may be too large to find an optimal potential which match best with the reference database. It should be noted that the transferability of a potential is reflected by the number of unconsidered physical quantities which could be reproduced, rather than the number of properties incorporated in the construction phase. To control the increment of FPs while keep the precisions of potentials, it may be helpful to employ parts of the target reference data, while the left reference data are served as a comparison with results extrapolated by the potentials. For brevity, we term the smaller reference data set as "SRD". Obviously, sensitivities of extrapolated results to uncertainties of the SRD are main obstacles to obtain a transferable potential. Recently, studies [26, 27] on the sensitivities of modified EAM potential of aluminum to the input uncertainties show that all modified EAM parameters interdependently influence the model outputs to varying degrees. That is to say, extrapolated results are insensitive to some FPs which may be not necessary for the potential. Unfortunately, the methods of reducing the redundant number of FPs are not specially discussed. Here, we find that the knowledge of SRD and its influence on the extrapolated results could be employed to reducing the redundant number of FPs. Besides, the FP number required for describing a certain system is largely determined by reliabilities of potential models or the functional forms adopted by the model. Some analytic style EAM potentials, represented by the works of Johnson [28, 29], are proposed and successfully applied to model complex metallic systems [30]. Unlike many of existing EAM models [17, 19, 31, 32] where the universal equation of states [33] (or Rose equation ) is served as a basic equation, the equation of states (EOS) is an extrapolated results of the analytic EAM potential of Johnson. Their embedding term is a universal function which is widely adopted in the constructions of EAM potentials. Though the extrapolated EOS at high pressure is not always satisfactory, the analytic scheme could effectively reduce the FP number and promote efficiencies of constructions. A drawback of the analytic EAM, as well as its variations (such as modified analyses EAM potential [12]), is that it may not have sufficient flexibilities to satisfy all the quantities of interest, and eventually affect its precisions when applied to complex cases which involve a large target database set needed to be reproduced. Hence, a suitable analytic-numerical approach may be expected to hold both advantages of the cubic-spline style potentials and the analytic potentials, and thus to reduce the FP number of potentials without affecting the precisions.

To do this, we propose a new EAM model based on approximations of the $p$-th moment to tight binding theory. To compensate the discrepancy of energy, as well as other physical quantities, predicted by the $p$-th moment approximations, a pairwise style interaction is employed in this model which could be thus determined in a direct manner from physical quantities in aids of numerical interpolation schemes. Details of the new model are described in part II. Then in part III, the new EAM model is validated by constructing and testing three potentials of aluminum. It is known that aluminum has a relatively accurate database due to its absence of $d$ electrons, and could be well simulated without angular force. These features could help us avoid the possible difficulties brought by uncertainties of the reference database and spherically averaged approximations adopted in the EAM framework. Our results show that the new potential has fewer FP number and smaller cutoff distance than some commonly used potentials of aluminum. Additionally, the equivalence between our new manybody term and the universal function at vicinities of equilibrium states are addressed, which may be helpful for understanding the physical bases of EAM models constructed via the universal function.

## II.  THEORETICAL METHOD

### A.  EAM Potential

According to the formulism of EAM potential [14], total energy of any configuration of nuclei can be expressed as a summation of atomic energy, that is

$$E = \sum_i \sum_j \phi(r_{ij}) + \sum_i F(\rho_i), \qquad (1)$$

where $\phi(r_{ij})$ represents pairwise interaction potential between atom $i$ and neighbor atom $j$ at a radial distance $r_{ij}$, and $F(\rho_i)$ is the so-called embedding energy which is a nonlinear function of total electron density $\rho_i$ at site $i$. The embedding energy-electron density curves, for H through Ar, have been calculated within a density-functional scheme [34, 35]. Banerjea et al [36] further express the scaled embedding energy-electron density curves as a universal function which could be parameterized as

$$F(\rho_i) = -F_0 \left[1 - n \ln\left(\frac{\rho_i}{\rho_e}\right)\right]\left(\frac{\rho_i}{\rho_e}\right)^n, \qquad (2)$$

where $F_0$ and $n$ are parameters. $\rho_e$ is the electron density at equilibrium state (where total energy of the lattice reaches a minimum with respect to the change of lattice parameter). If not specified, we will use subscript $e$ to discriminate the quantities of equilibrium state from others and neglect the subscript of the central atom index. Although expression (2) is physically reasonable at large $\rho_i$, it approaches zero at $\rho_i = 0$ with infinite slope and yields poor values for the pressure derivative of bulk modulus [28]. Wadley avoided the problem via using expression (2) only at large $\rho_i$, while the remaining segments of embedding function is replaced by simple cubic polynomials [28, 37]. Below, we will interpret the embedding energy (i.e., manybody term) in a different way, which gives a new EAM approach to interatomic potentials.

Inspired by the work of Finnis and Sinclair [16], the total energy of a system of interest could be evaluated through the $p$-th moment approximation of tight binding theory [38-40] which is essentially a manybody interaction. The $p$-the moment ($\mu_p$) of the density of states connects to band energy, whose center has been shifted to zero, and Hamiltonian matrix $\boldsymbol{H}$ by

$$\mu_p = \int E^p n(E) dE = \text{Tr}[\boldsymbol{H}^p] = \sum_i [\boldsymbol{H}^p]_{ii}. \qquad (3)$$

where $n(E)$ is local density of states. In tight binding theory, the diagonal and off-diagonal terms of N-dimension matrix $\boldsymbol{H}$ are corresponding to onsite and hopping integrals for N-atom cluster, respectively. Because $[\boldsymbol{H}^p]_{ii}$ involves $p$-hop chains (that is $H_{ij}H_{jk}H_{kl}\ldots H_{mi}$) and could be calculated from local topology, the $p$-the moment reflects $p$-body interactions. For simple metals where atom $i$ and its surrounding atoms are equivalence, $\sum_i [\boldsymbol{H}^p]_{ii}$ could be represented by $\rho_i$ where $\rho_i \equiv \mu_p$ is proportional to $\sum_j b(r_{ij})^\theta$ and $b(r_{ij})$ is a two-center hopping integral, which therefore stands for the contribution of atom $j$ to $\rho_i$. In above derivations, we have assumed that the $p$-hop chain of $[\boldsymbol{H}^p]_{ii}$ could be expressed by exponentiation of two-center hopping integral. Except for interaction models of the first nearest neighbors, the exponent is not necessary equal to $p$ for the models whose interaction range is beyond the first nearest neighbors, which has been denoted by $\theta$ instead. In addition, from the first equation of (3), the energy $E$ relates to $\mu_p$, thus $\rho_i$, by $E \propto \sqrt[p]{\rho_i}$ for a rectangular band. Assuming that the pairwise interaction potential in expression (1) serves as a correction to energetic, structural and elastic properties of a certain material, the energy *E* is corresponding to the embedding energy in equation (1). This idea is equivalent to the tight binding theory [41] if *E* is corresponding to valence bond and promotion energies, and the pairwise interaction is corresponding to total electrostatic and exchange-correlation energies. Further, by considering transformation invariance relations of EAM potential and setting the first derivative of embedding energy at $\rho_e$ to be zero like Banerjea [36] do, we have

$$F(\rho_i) = F_0'\left(p\sqrt[p]{\rho_i/\rho_e} - \rho_i/\rho_e\right), \tag{4}$$

where $p = 1/n^2$, $F_0' = F_0/(1-p)$. In our construction procedures, *p* is allowed to be a fractional number for convenience. Thus, *p* stands for an "effective" order of the moments. It is worth noting that the meaning of $\rho_i$ in the above expression is different from the one in expression (2), which will be described further below. From the above derivation, expression (1) is still valid within the new framework except for the physical meanings of the embedding term and the pairwise interaction. That is to say, the ideas of the new approach have no relations with the original EAM model, which is termed to be "EAM" simply because of their similar functional form of the total energy. We will use expression (2) at $\rho_i \geq \rho_e$ and expression (4) at $\rho_i < \rho_e$ for the construction of the first EAM potential (EAM-I) of aluminum, and only use expression (4) for the construction of the second one (EAM-II). The results of these two potentials, corresponding to two different EAM models, will discussed in part III. In the original EAM model, the total electron density of site *i* is assumed to be a linear superposition of electron density per atom surrounding the site. A spherical atomic electron density ($f(r_{ij})$) is found to be good enough for many metallic systems [12, 42], but poorly satisfied for covalently bonded systems where angular dependency cannot be neglected [43, 44]. In present work, the former is our major consideration while the later will left for further works. Then, $\rho_i$ is given by

$$\rho_i = \sum_{j \neq i} f(r_{ij}), \tag{5}$$

where the summation of the atomic electron density is over all neighbors defined by a cutoff distance $r_{ce}$. The spherical atomic electron density has been assumed to be several kinds of analytic forms semi-/empirically [12, 20, 22, 30]. According to the derivation of expression (4), the definition of expression (5) is still valid for $\rho_i$ in the expression (4) when $f(r_{ij}) \propto b(r_{ij})^\theta$. The expression of $f(r_{ij})$ in our previous work [12] is found to meet this requirements, that is

$$f(r_{ij}) = f_0 \left(\frac{r_{1e}}{r_{ij}}\right)^\theta \left(\frac{r_{ce}-r_{ij}}{r_{ce}-r_{1e}}\right)^2, \tag{6}$$

where *θ* is a potential parameter. This means that $b(r_{ij})$ takes the form of $r_{1e}/r_{ij}$. The value of $f_0$ has no effect on calculations of elemental qualities, which is retained for constructions of alloy potential in the future. Then, parameterization procedures could begin if an expression of the pairwise interactions is given. However, rather than adopting this routine, we take an analytic scheme, which cooperated with the weighted sum method, to determine the pairwise interaction dynamically. In the next section, the analytic scheme is introduced to find pairwise interactions dynamically, which could effectively avoid the theoretical difficulties from the lack of sufficient knowledge of pairwise interactions.

## B. An Analytic Scheme of Pairwise-Interaction Construction

The precise solution of pairwise interactions associates with an extremely complex ion-electron system governed by quantum mechanics, which are usually unknown except for some simple case (such as hydrogen). Alternatively, the pairwise interactions could be inferred from physical quantities of elements. One typical example is the applications of lattice inversion method [45-47] to calculate interatomic potentials from equation of states. Here, we calculate the pairwise interactions from some constantly considered physical quantities of elements which include lattice parameter, cohesive energy, single vacancy formation energy and elastic constants. The calculation procedure is different from the traditional analytic construction scheme because decoupling between many body effects, denoted by the embedding term, and pairwise interactions is considered.

Before introduce the analytic scheme, we recall some basic properties of a crystal. The first one is cohesive energy which could be expressed by

$$E_{coh} = \frac{1}{2}\sum_{i \neq j} \phi(r_{ij}) + F(\rho_e) = \frac{1}{2}\sum_I n_I \phi(b_I a_0) + F(\rho_e), \tag{7}$$

where $a_0$ is lattice parameter at equilibrium states. $n_I$ denotes atom number of the $I$-th nearest neighbors and $b_I$ is the corresponding separation distance reduced by $a_0$, which could be evaluated by minimizing the total energy of different lattice parameters. That is to say, the derivative of the total energy with respect to lattice parameter is zero at $a_0$. Then we have

$$E'(a_0) = \frac{1}{2}\sum_I n_I b_I \phi'(b_I a_0) + F'(\rho_e) \sum_I n_I b_I f'(b_I a_0) = 0. \tag{8}$$

The formation energy of single vacancy is usually concerned in atomic simulations because it is the simplest point defect whose un-relaxed formation energy is about the order of ~ 0.1eV larger than relaxed one. Supposing that a vacancy is introduced into an ideal crystal, originally consisting of $\mathcal{N}$ atoms with cohesive energy of $E_{coh}$ per atom (i.e., its total energy $E_\mathcal{N}$ is $\mathcal{N}E_{coh}$.), the total energy of the resulting system, containing one vacancy, is $E_{\mathcal{N}-1}$. Then, the un-relaxed vacancy formation energy is given by

$$E_{1v} = E_{\mathcal{N}-1} - (\mathcal{N}-1)E_{coh} = (E_{\mathcal{N}-1} - E_\mathcal{N}) + E_{coh} = \Delta F - E_{coh} + F(\rho_e), \tag{9}$$

where

$$\Delta F = \sum_I n_I \left[ F(\rho_e - f(b_I a_0)) - F(\rho_e) \right]. \tag{10}$$

To involve the second derivatives of EAM potentials, elastic properties of a crystal are often concerned. From lattice dynamics theory, elastic constants of a pure element could be expresses as

$$C_{\alpha\beta\mu\nu} = -\frac{1}{2\Omega_0} \sum_{k,k'} \Phi_{\alpha\beta}(k,k') r^\mu_{kk'} r^\nu_{kk'}, \tag{11}$$

where $\Omega_0$ is the volume per atom of the crystal at equilibrium states, and α, β, μ and ν denote Cartesian index. The summation, with respect to atom $k$ and $k'$, runs over all atoms in the system of interest. Force constant $\Phi_{\alpha\beta}(k,k')$ could be evaluated by

$$\Phi_{\alpha\beta}(k,k') = \left[-\phi''(r_{kk'}) + \frac{\phi'(r_{kk'})}{r_{kk'}}\right] \frac{r^\alpha_{kk'} r^\beta_{kk'}}{r^2_{kk'}} - \frac{\phi'(r_{kk'})}{r_{kk'}} \delta(\alpha - \beta) - (F'_k + F'_{k'}) \left\{\left[f''(r_{kk'}) - \frac{f'(r_{kk'})}{r_{kk'}}\right] \frac{r^\alpha_{kk'} r^\beta_{kk'}}{r^2_{kk'}} + \frac{f'(r_{kk'})}{r_{kk'}} \delta(\alpha - \beta)\right\} + F''_{k'} f'(r_{kk'}) \frac{r^\alpha_{kk'}}{r_{kk'}} \sum_{i \neq k'} f'(r_{k'i}) \frac{r^\beta_{k'i}}{r_{k'i}} - F''_k f'(r_{kk'}) \frac{r^\beta_{kk'}}{r_{kk'}} \sum_{i \neq k} f'(r_{ki}) \frac{r^\alpha_{ki}}{r_{ki}} + \sum_{i=k,k'} F''_i f'(r_{ki}) f'(r_{k'i}) \frac{r^\alpha_{ki} r^\beta_{k'i}}{r_{ki} r_{k'i}}. \tag{12}$$

From formula (2) or (4), we have

$$F(\rho_e) = F_0, \quad F'(\rho_e) = 0, \quad F''(\rho_e) = F_0 n^2 / \rho_e^2. \tag{13}$$

Then, formula (8) and (11) could be rewritten as

$$E'(a_0) = \frac{1}{2}\sum_I n_I b_I \phi'(b_I a_0) = 0 \tag{14}$$

and

$$\Omega_0 C_{\alpha\beta\mu\nu} = B_{\alpha\beta\mu\nu} + F''(\rho_e)V_{\alpha\beta}V_{\mu\nu}, \tag{15}$$

where

$$B_{\alpha\beta\mu\nu} = \frac{1}{2}\sum_i \left[-\phi''(r_{ie}) + \frac{\phi'(r_{ie})}{r_{ie}}\right] r_{ie}^\alpha r_{ie}^\beta r_{ie}^\mu r_{ie}^\nu / r_{ie}^2, \tag{16}$$

$$V_{\alpha\beta} = \sum_i \rho'_{ie} r_{ie}^\alpha r_{ie}^\beta / r_{ie}. \tag{17}$$

For cubic lattice, three independent elastic constants in Voigt notation are

$$C_{11} = [B_{11} + F''(\rho_e)V_{11}^2]/\Omega_0, \tag{18}$$

$$C_{12} = [B_{12} + F''(\rho_e)V_{11}^2]/\Omega_0, \tag{19}$$

and

$$C_{44} = B_{12}/\Omega_0. \tag{20}$$

The effects of pairwise interaction and manybody term on the elastic properties could be seen more clearly by rewriting formula (18) and (19) in the bellowing forms:

$$C' = \frac{1}{2}(C_{11} - C_{12}) = \frac{1}{2}(B_{11} - B_{12})/\Omega_0, \tag{21}$$

$$C_d = C_{12} - C_{44} = F''(\rho_e)V_{11}^2/\Omega_0. \tag{22}$$

That is to say, $C'$ and $C_{44}$, as importance indicators of lattice stabilities, are completely determined by pairwise interactions, while $C_d$ is uniquely determined by the embedding term. From the discussions above, we find that equation (14), (19) and (20) only depend on $\phi(r_{ij})$ and its derivatives, while equation (7) and (9) rely on both pairwise interaction and embedding term. Namely, the coupling effects between manybody term and pairwise interactions are only reflective in cohesive energy and single vacancy formation energy in present situation. Therefore, $\phi(r_{ij})$ could be evaluated at certain points via equation (7), (9), (14), (19) and (20) if all parameters of embedding term, that are $F_0$, $n$, are known. To this end, multi-objective optimization (MOO) procedure [17, 22] is employed to achieve the search of the embedding parameters as well as other undetermined parameters of pairwise interactions, where the remaining parameters of $\phi(r_{ij})$ could be calculated in every try of the search. To incorporate the information between each equilibrium separation and obtain a reliable potential via this scheme, a robust numerical scheme for solving equations of $\phi(r_{ij})$ is needed, which will be described in the next section.

### C. Numerical Scheme and Parameterization of EAM Potential

From the analyses above, the basic qualities at equilibrium states could be guaranteed if equation (7), (9), (14), and (20) - (22) are well satisfied. Among these relations, only equation (7), (14), (20) and (21) are employed for solving the value of $\phi(r_I)$ as well as its derivatives at each equilibrium separations $r_I$. While equations (9) and (22), associated with single vacancy formation energy and Chauchy pressure, are served as target qualities involved in the MOO procedure. It should be noted that the four equations cannot be determine all values of $\phi(r_I), \phi'(r_I)$ and $\phi''(r_I)$, where $I = 1, 2, …, N$ if we take the cutoff distance $r_c$ of $\phi(r)$ to be

$$r_c = r_N + k_c(r_{N+1} - r_N), \tag{23}$$

where $k_c$ is a potential parameters ranging from 0.1 to 0.9. That is to say, to completely determine the values of $\phi(r_I)$ and its derivatives, we still lack $3N - 4$ equations. Besides, values of $\phi(r)$ between two adjacent equilibrium separations cannot be reflected by the qualities at equilibrium states. In order to solve these two problems, we firstly find an approximate numerical solution of $\phi(r)$ at equilibrium states, and then correct it with other qualities of both equilibrium and non-equilibrium states. In present work, we start by assuming $\phi''(r)$ to be nearly linearly changed between adjacent $r_I$ and thus linear interpolation could be used between adjacent $\phi''(r_I)$, which will result in a cubic spline interpolation between adjacent $\phi(r_I)$. If we select $\phi(r_I)$ and $\phi''(r_I)$ as parameters to be determined, that is

$$\phi(r_I) = y_I, \phi''(r_I) = M_I, \quad (I = 1, 2, \ldots, N + 1), \tag{24}$$

where we have used the notation $r_{N+1}$, $y_{N+1}$ and $M_{N+1}$ to denote $r_c$, $\phi(r_c)$ and $\phi''(r_c)$, respectively. If not specified, these notations will be held in the rest of this section. Then the expression of $\phi(r)$ could be obtained via integrating twice of $\phi''(r)$, that is

$$\phi(r) = \frac{(r_{I+1}-r)^3}{6h_I} M_I + \frac{(r-r_I)^3}{6h_I} M_{I+1} + \left(y_I - \frac{h_I^2}{6} M_I\right)\frac{r_{I+1}-r}{h_I} + \left(y_{I+1} - \frac{h_I^2}{6} M_{I+1}\right)\frac{r-r_I}{h_I}, \quad (r_I \leq r < r_{I+1}, I = 1, 2, \ldots, N), \tag{25}$$

where

$$h_I = r_{I+1} - r_I. \tag{26}$$

The derivation of equation (25) has already considered the continuity conditions of $\phi(r)$ and $\phi''(r)$ at $r_I$ ($I = 2\ldots, N$), while the continuity conditions of $\phi'(r)$ should be considered separately, which could be written as

$$\phi'(r_I + 0) = \phi'(r_I - 0), \quad (I = 2, \ldots, N). \tag{27}$$

Considering equation (24) and (25), equation (9) and (14) could be expressed as

$$\sum_{I=1}^{N} n_I y_I = 2(E_{coh} - F_0), \tag{28}$$

$$n_1 b_1 h_1 M_1 + \sum_{I=2}^{N} \left(\frac{1}{2} n_{I-1} b_{I-1} h_{I-1} + n_I b_I h_I\right) M_I$$

$$+ 3\left[\frac{n_1 b_1}{h_1} y_1 + \sum_{I=2}^{N} \left(\frac{n_I b_I}{h_I} - \frac{n_{I-1} b_{I-1}}{h_{I-1}}\right) y_I\right] = 0, \tag{29}$$

where $b_I$ relates to $r_I$ by

$$b_I a_0 = r_I. \tag{30}$$

Similarly, equation (20) and (21) could be rewritten as

$$a_1 y_1 - \sum_{I=2}^{N}(a_{I-1} - a_I)y_I + A_1\left(1 + \frac{1}{3}h_1'\right)M_1 + \sum_{I=2}^{N}\left[\frac{1}{6}A_{I-1}h_{I-1}' + A_I\left(1 + \frac{1}{3}h_I'\right)\right]M_I = D, \tag{31}$$

where

$$\begin{cases} A_I = \sum_{\{r_I\}} r_{I,x}^2 r_{I,y}^2 / r_I^2 \\ D = 4\Omega_0 C_{44} \end{cases}, \tag{32}$$

for equation (20) and

$$\begin{cases} A_I = \sum_{\{r_I\}} \left(r_{I,x}^4 - r_{I,x}^2 r_{I,y}^2\right)/r_I^2 \\ D = 4\Omega_0 C' \end{cases} \tag{33}$$

for equation (21), respectively. $h_I'$ and $a_I$ is defined as

$$h_I' = h_I/r_I, \tag{34}$$

and

$$a_I = \frac{A_I}{r_I h_I}. \tag{35}$$

This numerical scheme contains totally $N+3$ equations of $2(N+1)$ parameters, that is $\{y_I, M_I\}$ ($I = 1, 2, \ldots, N, N+1$). Below, we will use $\{y_I, M_I\}$ to denote the unknown parameters of pairwise interactions. In comparison with the $2(N+1)$ unknown parameters, $N-1$ more equations are required to form complete equations. Here, the third neighbor model (3NM) will be developed in details, whose idea could be easily generalized to the N-th neighbor model by introducing ($N$-3) additional parameters. For 3NM, two more equations are required. It is naturally thought that equation (9) could serve as one equation for 3NM. However, we find that equation (7) and (9) cannot be included in the equation sets simultaneously, otherwise the numerical solution of $\{y_I, M_I\}$ may turn out to be infinite. To avoid this problem, we set $y_{N+1}$ and $M_{N+1}$ to be zero instead of utilizing equation (7) to solve $\{y_I, M_I\}$. Because expression (25) does not guarantee that the first order derivative of $\phi(r)$ at $r_c$ is zero, $\phi(r)$ is redefined at the range of $[r_N, r_c]$ (or $[r_N, r_{N+1}]$) below:

$$\phi(r) = \left[\frac{(r_{N+1}-r)^3}{6h_N}M_I + \frac{(r-r_N)^3}{6h_N}M_{N+1} + \left(y_N - \frac{h_N^2}{6}M_N\right)\frac{r_{N+1}-r}{h_N} + \left(y_{N+1} - \frac{h_N^2}{6}M_{N+1}\right)\frac{r-r_N}{h_N}\right]h(r), \quad (36)$$

where tailor function is

$$h(r) = \sum_{i=0}^{5} \lambda_i \left(1 - \frac{r_c - r}{r_c - r_N}\right)^i, \quad (37)$$

where $\lambda_i$ take values which makes $h(r)$ satisfy

$$\begin{cases} h(r_N) = 1, & h'(r_N) = 0, & h''(r_N) = 0 \\ h(r_c) = 0, & h'(r_c) = 0, & h''(r_c) = 0 \end{cases}. \quad (38)$$

Therefore, the pairwise interaction could be uniquely determined via values of cohesive energy, lattice parameter and elastic constants measured at equilibrium states if $F_0$, as a coupling parameter, is known. Here, we have used the term "coupling parameters" to represent undetermined parameters (or potential parameters) of our potential in order to emphasis its coupling nature between pairwise interaction and embedding energy. In other words, any changes of the coupling parameters will affect energy partitions between pairwise interaction and embedding energy, which could be determined by an optimization procedure. Unfortunately, the resulting potential of this scheme fails to reproducing dispersion relations and single vacancy formation energy simultaneously, no matter how complex of atomic electron density function is used. The main reason is that the interpolation scheme is not precise enough to calculate $\phi'(r_I)$ despite of the accuracy of $\phi(r_I)$ and $\phi''(r_I)$. To improve the calculation precision of $\phi'(r_I)$, Z additional nodes are inserted into original nodes defined at equilibrium separations. Still, continuity conditions are obeyed at all nodes, which contributes to Z additional equations while introduce 2Z unknown parameters $\{y_i^a, M_i^a\}$ ($i = 1, 2, \ldots, Z$). Again, the remaining Z parameters $\{M_i^a\}$ are taken to be coupling parameters which will be optimized by fitting them to other physical quantities through MOO method. Thus, there are $Z+4$ potential parameters to be optimized, which are $n$, $F_0$, $\theta$, $k_c$, $\{M_i^a\}$ ($i = 1, 2, \ldots, Z$). The additional nodes make the potential more flexible and enable the potential to adapt various application circumstances.

Additionally, qualities at extreme conditions may be involved in the fitting procedures which requires a complete definition of $\phi(r)$. We use Morse function, consisted of a repulsive and an attractive term, as short range pairwise interactions, that is

$$\phi(r) = D_e\left(e^{-2\alpha(r-r_e)} - 2e^{-\alpha(r-r_e)}\right), \quad (r < r_0) \quad (39)$$

where $r_0 = \chi r_1$, $0 \leq \chi \leq 1$. The value of $\chi$ ranges from 0.8 to 1.0 depending on elements. Other commonly used two-body functions, like LJ function, could also be applied here. Though it may affect the equation of states at compressed states, the quality of the potential is unchanged for applications at near equilibrium states based on our testing. The remaining parameters ($D_e$, $\alpha$, $r_e$) in expression (39) could be determined by smooth connection conditions at $\phi(r_0)$. The cutoff distance of atomic electron density is $r_{ce} = r_{N+1} + 0.5(r_{N+2} - r_{N+1})$, where $r_{N+1}$ and $r_{N+2}$ are the (N+1)-th and (N+2)-th nearest neighbor separation of the reference lattice at equilibrium states, respectively.

In summary, this potential model contains only five FPs which are $n$, $F_0$, $\theta$, $k_c$ and $\chi$. Several additional FPs could be involved in the model through additional nodes inserted between equilibrium separations depending on application circumstances. The additional FP number is equal to the number of additional nodes for 3NM. To make the total FPs as small as possible, we control the number of additional nodes employed to reproduce the SRD of a target application. However, the final FP number relies on the reliabilities of a potential model for a certain system. In the next part, we will test this model in aluminum.

## III. RESULTS AND DISCUSSIONS

### A. Fitting Parameters of Interatomic Potentials of Aluminum

The method mentioned in part II has been implemented in a software tool, called "Constraint Mult-Object Fitting Proedure" (CMOFP), by us. CMOFP utilizes the MOO method combined with some optimization algorithms (the simplex algorithm of Nelder and Mead is employed here) to find the optimal potential parameters which achieves a best match between predicted results of the new potential and reference database. In present work, we intend to construct a 3NM potential of aluminum with smaller FP number than some reported potentials [22, 30, 48] while still as accurate as them. Our reference database consists of lattice parameter, cohesive energy, single vacancy formation energy, elastic constants, phonon dispersion spectra and equation of states at stretched states. An extended reference database is used to construct 3NM potential EAM-I and EAM-II, which include stacking faults energy (SFE) and unstable stacking fault energy (USFE) except for the quantities contained in the basic reference database. To find the SRD, the basic reference database is also used to construct 3NM potential EAM-II' as a comparison with EAM-II. EAM-I and EAM-II are different in their functional form of embedding energy as mentioned above, while the embedding function of EAM-II' is the same as EAM-II. Two additional nodes are inserted between the first and the second nearest neighbor separation, and one additional node is inserted between the second and the third nearest neighbor separation. After several trials, the resulting potentials are insensitive to the detailed positions of these three additional nodes which are simply fixed at $r_1^a = 3.1$, $r_2^a = 3.5$, $r_3^a = 4.3$ and $r_4^a = 4.6$. With these settings, optimal parameters ($n$, $F_0$, $\theta$, $k_c$, $M_1^a$, $M_2^a$, $M_3^a$, $M_4^a$) of the three potentials are obtained by the CMOFP and listed in Table I. As shown in Table I, the number of aluminum potential parameters is 9, while it is 20-40 for the force-matching approach [19], 16 for MEAM [49] and 23 for a recently reported Charge optimized manybody potential for aluminum [48]. Compared with these potentials, much smaller FP number is involved in our potentials. Fig.1-3 have shown the EAM functions of the three potential. Obviously, embedding function defined by expression (4) is stiffer than the one defined by (2) at large $\rho$, which will be discussed further in the next section.

### B. Testing the Potentials of Aluminum

Critical properties of aluminum are predicted by the three potentials as well as two commonly used potentials developed by Zhou et al. [30] and Mishin et al. [22] (See Table II). Our results show that intrinsic stacking fault energy $\gamma_{SF}$<112> cannot be correctly predicted without incorporating it into the reference database, which is an important property for descriptions of mechanical behaviors of aluminum. Other unfitted qualities, that is cohesive energy of BCC and HCP structures and surface energies, could also reasonably agree with experimental measurements or ab initio calculations when the best fitting to the reference database is achieved. The best fitting means that the value of multiple-objective function reaches its minima, which could be achieved by repeating the fitting procedure three or four times. The good agreements between the prediction results (in Table II and Fig. 4) and the reference data indicate that we have achieved the best fitting with the potential parameters. Although the cohesive energies of BCC and HCP structure are bigger than that of FCC structure as expected, the cohesive energy of BCC structure changes approximately from -3.34 to -3.10 depending on the position of additional nodes, while the cohesive energy of HCP structure will always located at around of the correct value (-3.33 eV) when the best fitting is arrived. After including the SFE into the basic reference database and confining USFE to be 0.8 eV larger than the SFE with a small weight, the cohesive energy of BCC structure could be effectively converged to the vicinity of the correct value (-3.25 eV). That is to say, the extended reference database is our SRD in present studies.

Equation of states is of fundamental importance for describing mechanical response to shock compressions [50, 51]. As shown in Fig. 5, equation of states of aluminum at compressed states has been predicted at 0 K with the three potentials. The results of Zhou el al. and Mishin el al. are also provided in the figure. From the results, we find that the equation of states predicted by us agrees quite well with experiment measurements [52, 53] up to more than 150 GPa just as the results of Mishin el al. In spite of the similar results, the difference between ours and Mishin's lies at the construction procedures. Mishin et al.[22] have utilized the equation of states, in terms of the universal Rose equation, as one of the function to construct their potential of aluminum which will surely result in the correct equation of states. While the equation of states at compressed states is not incorporated in our reference database which is merely a result of extrapolations from equilibrium and stretched states to the compressed states. We find that more FPs are needed to match the reference database if the universal Rose equation instead of equation (4) is present in our EAM potentials. Furthermore, the predicted equation of states at compressed states is sensitive to the uncertainties of reference database, which cannot be studied by potentials containing the universal Rose equation. We will discuss the roles of the sensitivity played on improvements of transferability without adding additional FPs in the next section. Besides, as shown in Fig. 2, embedding function defined by (4) could generate similar behaviors at vicinities of $\rho_e$ as the one defined by expression (2). And the precisions of EAM-I and EAM-II, corresponding to that of (4) and (2) at $\rho > \rho_e$, are also comparable (See Table II and Fig. 4-5). Then expression (4) may provide another explanations for (2) whose the empirical parameter $n$ could relate to the effective order of moments of local density of states by $p = 1/n^2$. If the order of moments is much larger than two, our approximations in the derivations of expression (4) should be improved in order to make a satisfactory prediction. Considering the similar behaviors between expression (4) and (2), this condition may be also applied for (2). Fortunately, the effective orders of moments for aluminum under the concerned cases are within ranges of (1.8, 1.98), which meets the condition. This may be a reason that the new EAM model could correctly simulate the bulk properties of aluminum and exhibit a good compatibility between the SRD and the extrapolated results (See section C).

Melting point is a critical property for crystal materials at elevated temperature, which will be predicted by molecular dynamics (MD) simulations using the three new potentials of aluminum. To observe the melting of aluminum, a $30a_0 \times 30a_0 \times 30a_0$ single crystalline aluminum sample is equilibrated over the range from 600 K to 1500 K at 1 atm via constant temperature and pressure simulations for 80 ps, respectively. Each simulation is performed with Large-scale Atomic/Molecular Massively Parallel Simulator[54]. Nosé–Hoover thermostat and barostat are employed to fix temperature and pressure during the simulations. Bond order parameters[55] are employed to distinguish isentropic liquids from crystalline solids, which are calculated and averaged over the last one thousand steps for each MD simulations. Parameter $Q_6$ as a function of temperature ($T$) is shown in Fig. 6 where the sudden drop in the value of $Q_6$ indicates the occurrence of melting. Thus, the predicted melting point is about 1200K, 1100K and 1100K for EAM-I, EAM-II and EAM-II', respectively, which are larger than the experimental value 933K [56]. Because the aluminum sample employed here is an ideal single crystal whose melting point will decrease in the presence of lattice defects, for example dislocations and grain boundaries. Besides, this prediction method of melting point is the so-called gradually heating method, whose results are usually larger than its real ones because of extraordinarily high heating rate [57]. More reliable prediction could be made through free energy approach, which will be conducted in the next paragraph. The melting transition could also be observed through adaptive common neighbor analyses [58] (See Fig. 7). Additionally, there is a sudden rising in the $Q_6$-$T$ curve near the melting point for EAM-II', while the behaviors of aluminum do not have the abnormality for EAM-I and EAM-II. This results suggest that the extended reference database employed here does not affect the prediction of melting point, but do have an influence on the detailed material behaviors near the melting point. By comparing the melting point predicted by EAM-I and EAM-II, we find that embedding function defined by expression (4) gives more accurate melting point than that of expression (2).

Alternatively, the melting point could be predicted by equating the free energy of solid phase with that of liquid phase. As a complementation of the predictions of melting points by gradually heating method, the free energy approach is employed to calculate the melting point of aluminum with EAM-II. Here, free energies of solid and liquid phase are calculated by reversible scaling (RS) technique [59, 60] and adiabatic switch (AS) [60-62]. Supposing that the system of interest contains $N$ atoms of equal mass $m$, its Hamiltonian could be expressed as

$$H_0 = \sum_{i=1}^{N} \frac{p_i^2}{2m} + U_0(r_1, \dots, r_N), \tag{40}$$

where $r_i$ and $p_i$ are the position and momentum vectors of atom $i$, $U_0$ denotes potential energy. If the system is in thermal equilibrium under constant temperature $T_0$ and constant volume $V$, the Helmholtz free energy is given by

$$F_0(T_0) = -k_B T_0 \ln\left[\int d^{3N}r \exp\left(-\frac{U_0}{k_B T_0}\right)\right] + 3N k_B T_0 \ln \Lambda(T_0), \tag{41}$$

where $k_B$ is Boltzmann's constant and $\Lambda(T_0)$ denotes the thermal de Broglie wavelength $h/\sqrt{2\pi m k_B T_0}$. AS technique rescales the system of interest by a factor $\lambda$ ($\lambda > 0$) to create a new system whose Hamiltonian and Helmholtz free energy are written as

$$H_1(\lambda) = \sum_{i=1}^{N} \frac{p_i^2}{2m} + \lambda U_0(r_1, \dots, r_N) \tag{42}$$

and

$$F_1(T_0, \lambda) = -k_B T_0 \ln\left[\int d^{3N} r \exp\left(-\frac{U_0}{k_B T}\right)\right] + 3N k_B T_0 \ln \Lambda(T_0), \quad (43)$$

respectively, where $T = T_0/\lambda$. Considering equations (41) and (43), the temperature dependence of $F_0(T)$ could be expressed by the $\lambda$ dependence of $F_1(T_0, \lambda)$ at fixed temperature $T_0$,

$$\frac{F_0(T)}{T} = \frac{F_1(T_0, \lambda)}{T_0} + \frac{3}{2} N k_B \ln \frac{T_0}{T}. \quad (44)$$

If $\lambda$ changes its value from $\lambda(0)$ to $\lambda(t)$ with time $\tau$ slowly under constant $T_0$ and constant $V$ so that a reversible thermodynamic path is created between $\lambda(0)$ and $\lambda(t)$, then AS approach could be used to connect $F_1(T_0, \lambda(t))$ and $F_1(T_0, \lambda(0))$ via a cumulative reversible work $W(t)$,

$$\Delta F_1(\lambda(t), \lambda(0)) \triangleq F_1(T_0, \lambda(t)) - F_1(T_0, \lambda(0)) = \int_0^t d\tau \frac{d\lambda}{dt}\bigg|_\tau U_0(\boldsymbol{r}_1(\tau), \dots, \boldsymbol{r}_N(\tau)) \triangleq W(t). \quad (45)$$

Substituting equation (45) back to (44), the Helmholtz free energy at $T$ could be expressed as

$$\frac{F_0(T(t))}{T(t)} = \frac{F_0(T(0))}{T(0)} + \frac{W(t)}{T_0} - \frac{3}{2} N k_B \ln \frac{T(t)}{T(0)}, \quad (46)$$

where $T(t) = T(0)/\lambda(t)$, $T(0) = T(0)/\lambda(0)$. Thus, the temperature dependence of $F_0(T)$ could be determined by a time-dependence function $W(t)$ and its initial value $F_0(T(0))$ (or $F_0(T_0)$). Practically, the time $\tau$ dependence of $\lambda$ adopts the bellowing form

$$\lambda(\tau) = \frac{\lambda_0}{1+\frac{\tau}{t}\left(\frac{\lambda_0}{\lambda_t}-1\right)}, \quad (47)$$

where $\lambda_0 = \lambda(0), \lambda_t = \lambda(t)$. To obtain $W(t)$, we perform MD simulations under under constant $T_0$ and constant $V$ ensemble at zero pressure for FCC aluminum and its liquid phase, respectively. The corresponding settings are: $T_0 = 245$ K, $\lambda(0) = 1$ and $\lambda(t) = 0.204167$ (i.e., $T(t) = 1200$ K) for solid phase, and $T_0 = 1400$ K, $\lambda(0) = 1$ and $\lambda(t) = 1.55556$ (that is $T(t) = 900$ K) for liquid phase. Setting the switch time to 40 ps would meet the requirements of RS technique in this work. As shown in Fig. 8, $W(t)$ per atom are calculated by AS approach for solid and liquid phase of aluminum, respectively.

$F_0(T_0)$ is calculated by AS approach which could evaluate the difference of free energies between two systems via a composite Hamiltonian

$$H(\lambda) = K + \lambda U_1 + (1-\lambda)U_2, \quad (48)$$

where $K$ is kinetic energy, $U_1$ and $U_2$ are potential energy of the two systems, corresponding to $\lambda = 1$ and $\lambda = 0$, respectively. Under canonical ensemble, the Helmholtz free energy of the two systems are related by relation of

$$F(\lambda = 1) = F(\lambda = 0) + \int_0^1 \langle \frac{\partial H}{\partial \lambda} \rangle_\lambda d\lambda, \quad (49)$$

where $\langle \ \rangle_\lambda$ denotes ensemble average at certain $\lambda$. From equation (48), we have $\langle \frac{\partial H}{\partial \lambda} \rangle_\lambda = \langle U_1 - U_2 \rangle_\lambda$ whose integration over $\lambda$ is the so called reversible work. Therefore, $F(\lambda = 1)$ at certain temperature $T$ could be calculated by the reversible work if the free energy of the system corresponding to $\lambda = 0$ (reference system) is known. The reference system is taken to be Einstein solid for the calculations of solid phase. For convenience, we use $f$ to denotes the free energy per atom, that is

$$f(\lambda = 1) = f(\lambda = 0) + \frac{1}{N}\int_0^1 \langle U_1 - U_2 \rangle_\lambda d\lambda \triangleq f(\lambda = 0) + \Delta f. \quad (50)$$

The potential energy of Einstein solid is known to be

$$U_2 = \frac{1}{2} m \omega_E^2 \sum_i (\boldsymbol{r}_i - \boldsymbol{r}_{i0})^2, \tag{51}$$

Then the free energy is

$$f(\lambda = 0) = -3\left(1 - \frac{1}{N}\right) k_B T_0 \ln\left(\frac{T_0}{T_E}\right), \tag{52}$$

where Einstein temperature $T_E$ is

$$T_E = \hbar \omega_E / k_B. \tag{53}$$

Our simulation box consists of 10×10×10 primitive cells of FCC aluminum, totaling 4000 atoms. MD simulations, where λ ranges from 1 to 0 with an interval of 0.1, are run at 245 K for 40 ps at isobaric-isothermal condition with zero extra pressure. Einstein frequency $\omega_E$ is taken to be the value which makes the average root-mean-squared (rms) displacement equal to that of the original system, that is $m\omega_D^2$ = 3.219 eV/Å$^2$ ($m$ = 26.981 538 6 a.u.). Then the free energy of Einstein solid is $f(\lambda = 0)$ = 3.555 × 10$^{-3}$ eV. $\langle U_1 - U_2 \rangle_\lambda$ as a function of λ are shown in Fig. 9. Through numerical integration over λ, we obtain $\Delta f$ = -3.35066 eV. Thus the free energy of FCC aluminum at 245 K is $f_s(T(0)) = f(\lambda = 0) + \Delta f$ = -3.347105 eV.

Evaluation of $F_0(T_0)$ of the liquid phase is divided into two steps. In the first step, we connect the liquid phase of interest to a *weak* LJ liquid (whose attractive interactions between liquid atoms could be neglected) at constant $T_0$ and constant $V_0$ ensemble. The liquid phase consists of 4000 atoms which is obtained by directly heating FCC aluminum at $T_0$ = 1400 K with zero extra pressure. Interatomic potential of the LJ liquid is

$$U_l = 4\varepsilon \left[\left(\frac{\sigma}{r}\right)^{12} - \left(\frac{\sigma}{r}\right)^6\right], \; (r < r_c) \tag{54}$$

where $\varepsilon$ = 0.01 eV, $\sigma$ = 2.474873734 Å, $r_c = \sqrt{2}\sigma$ = 3.5 Å. According to the AS approach, the difference between these two systems is

$$\Delta f_1 = \frac{1}{N} \int_0^1 \langle U - U_l \rangle_\lambda \, d\lambda. \tag{55}$$

In the second step, the LJ liquid is expanded isothermally to the idea gas limit. During this process, the change of the free energy is

$$\Delta f_2 = \frac{1}{N} \int_{V_0}^\infty (P - P_I) \, dV, \tag{56}$$

where $P_I$ is the idea gas pressure. $\langle U - U_l \rangle_\lambda$ as a function of λ and $P - P_I$ as a function of $V$ have been shown in Fig. 10, respectively. According to equation (55) and (56), the free energy of the liquid phase at $T_0$ could be calculated by

$$f_l = \Delta f_1 + \Delta f_2 + f_{id}(T_0, V_0). \tag{57}$$

where $f_{id}$ is the free energy of idea gas. The values of each term in the above equation have been listed in Table III. Finally, the free energy of FCC aluminum and its liquid phase as a function of temperature at zero extra pressure are obtained as shown in Fig. 11. This approach predicts a bulk melting point of 982 K, comparable to the experimental result of 933 K [56]. The good consistence in bulk melting point between our result and the experimental data may result from the correct phonon dispersion relations predicted by our potential (EAM-II).

### C. Sensitivity of Extrapolated Results of Potentials to the Reference Data

According to our construction procedures, the qualities of constructed potentials are mainly influenced by the accuracies of the reference database. In this work, we could construct a transferable potential of aluminum partly because it has a basic reference database with small uncertainties. Among the basic reference database, the uncertainty of single vacancy formation energy is the largest which could reach as large as 10.96% [26]. Thus, it is necessary to discuss the sensitivity of extrapolated qualities of constructed potentials to the uncertainties of single vacancy formation energy. Here, we change the value of single vacancy formation energy in the reference database of EAM-II from 0.5 to 1.1 eV and reconstruct the potential EAM-II. Fig. 12 has shown the influence of single vacancy formation energy on the predicted EOS at compressed states, where we find that the predicted EOS would achieve a good match with experimental data when the value of single vacancy formation energy is within the range of (0.67, 0.80) eV. The value of the single vacancy formation energy by density functional theory is 0.72 eV, which is well located in this range. Moreover, the range of surface energies at (001), (110) and (111) with respect to the uncertainties of single vacancy formation energy are predicted to be (0.76, 0.87), (0.98, 1.12) and (0.68, 0.78) $J/m^2$, respectively (See Fig. 13). Values of the three surface energies calculated by density functional theory [48, 63] are 0.84, 0.91 and 0.75 $J/m^2$, respectively, while the experimental results are an averaged result of 0.98 $J/m^2$ over difference surfaces. This indicates that our predicted ranges of the surface energies agree well with the results of density functional theory or experiments. However, cohesive energy of HCP structure, as well as the corresponding lattice parameter, does not change with the increasing of the single vacancy formation energy, while cohesive energy and lattice parameter of BCC structure changes irregularly around -3.29 eV and 3.18 Å, respectively. It is due to the large distinctions between surroundings of atom in BCC and FCC structure so that the pairwise interactions at equilibrium separations of BCC structure cannot be evaluated with sufficient precisions through values at the equilibrium separations of FCC structure, which results in the uncertainties of the predicted cohesive energy of BCC structure. This could be improved by constraints of lattice parameter and cohesive energy of BCC structure. Similarly, due to the equilibrium separations of HCP structure are located nearby that of FCC structure, the pairwise interactions could be correctly evaluated at its equilibrium states and thus, we could obtain a correct cohesive energy of HCP structure. From the discussion above, we find that quantities of high precisions could be used to evaluate quantities of low precision by utilizing the sensitivities of extrapolated results to uncertainties of SRD. The sensitivities also provide an approach to match quantities of interest with reference data rather than adding additional FPs, which will eventually improve the precisions of the potential. It should be pointed that the correct extrapolations greatly rely on the reliabilities of potential model, which, in turn, proves the accuracies of our EAM model for describing properties of bulk aluminum.

## IV. SUMMARY AND CONCLUSIONS

In conclusion, we have discussed the methods of reducing FP number of EAM potentials from two aspects. Firstly, the FP number is controlled from the EAM model without affecting its flexibility and scalability. To do this, a new EAM framework is established, which is a generalization of the results of Finnis and Sinclair to the *p*-th moment approximations of tight binding theory. Under present EAM framework, pairwise interaction potential is employed to compensate the discrepancy of energy, as well as other physical quantities, predicted by the *p*-th moment approximations, which could be thereby evaluated through an analytic-numerical scheme. The new EAM model is equivalent to the tight binding model if the manybody term is viewed as valence bond and promotion energies, and the pairwise interaction is corresponding to total electrostatic and exchange-correlation energies. The analytic-numerical scheme, designed in this work, enables us to construct a potential without the need of knowing functional forms of the pairwise interactions in advance. And it could utilize several physical quantities (lattice parameter, cohesive energy, single vacancy formation energy, elastic constants, phonon dispersion frequencies and EOS) at equilibrium states and stacking fault energy, as well as the energy difference between the stacking fault and unstable stacking fault, to achieve a correct extrapolation to the compressed segments of pairwise interactions in the aids of Morse function. Other commonly used two-body functions can also be applied for the extrapolation, such as LJ function. Different extrapolation functions could generate a similar EOS up to 100 GPa without affecting physical quantities of elements at equilibrium states. The detailed numerical scheme for parameterizations of potentials of aluminum is a linear approximation of the second order derivatives of pairwise interactions, which results in a cubic spline function of the pairwise interactions. Due to the simple approximation, a few additional nodes are inserted between equilibrium neighbor separations to improve the precision of pairwise interactions at equilibrium neighbor separations, which is realized by the cooperation between the analytic scheme and the MOO procedure. Other numerical schemes of high precisions for calculating the pairwise interactions could also be implemented within the present framework. Validations of the potential model are demonstrated by constructions and testing of three aluminum potentials. Secondly, only a few additional FPs (or additional nodes for cubic spline) are selected to reproduce the SRD. The precisions of the potential are guaranteed via using the sensitivities of extrapolated results to the uncertainties of quantities in SRD. That is to say, the extrapolated results could be matched with the reference data through adjusting the values of quantities in SRD within the ranges of their uncertainties, but the effectivities of this method are dependent on the reliabilities of potential models adopted.

Besides, our results indicate that the new manybody term shows similar behaviors with a commonly used embedding function at the vicinities of equilibrium states, and the effective order ($p$) of moments of local density of states in our model relates to the empirical parameter $n$ in the latter one by $p = 1/n^2$. According to the derivations of the new manybody terms, we find that the new model could well reproduce bulk properties of simple metals if $p$ is not much larger than two. This condition may be also applied for EAM models using the latter embedding function.

# ACKNOWLEDGMENTS

The present work is supported by the National Natural Science Foundation of China (NSFC-NSAF U1530151 and NSFC 11102194, 11402243), National Key Laboratory Project of Shock Wave and Detonation Physics (No. 077120), the Science and Technology Foundation of National Key Laboratory of Shock Wave and Detonation Physics (Nos. 9140C670201110C6704 and 9140C6702011103) and Chinese National Fusion Project for ITER with Grant No. 2013GB114001.

# Reference

[1] S.K.R.S. Sankaranarayanan, V.R. Bhethanabotla, B. Joseph 2005 *Physical Review B* 71 195415.
[2] J. Wereszczynski, J.A. McCammon 2012 *Quarterly reviews of biophysics* 45 1.
[3] A.B. Belonoshko, N.V. Skorodumova, A. Rosengren, B. Johansson 2008 *Science* 319 797.
[4] N.Q. Vo, R.S. Averback, P. Bellon, S. Odunuga, A. Caro 2008 *Physical Review B* 77 134108.
[5] Z.T. Trautt, Y. Mishin 2012 *Acta Materialia* 60 2407.
[6] J. Schäfer, K. Albe 2012 *Acta Materialia* 60 6076.
[7] J.-L. Shao, P. Wang, A.-M. He, S.-Q. Duan, C.-S. Qin 2013 *Journal of Applied Physics* 113 163507.
[8] Y. Liao, M. Xiang, X. Zeng, J. Chen 2015 *Mechanics of Materials* 84 12.
[9] W. Ma, W.J. Zhu, Y. Hou 2013 *Journal of Applied Physics* 114 163504.
[10] J.A. Zimmerman, J.M. Winey, Y.M. Gupta 2011 *Physical Review B* 83 184113.
[11] W. Ma, W. Zhu, F. Jing 2010 *Applied Physics Letters* 97 121903.
[12] K. Wang, S. Xiao, H. Deng, W. Zhu, W. Hu 2014 *International Journal of Plasticity* 59 180.
[13] K. Kadau, T.C. Germann, P.S. Lomdahl, B.L. Holian 2002 *Science* 296 1681.
[14] M.S. Daw, M.I. Baskes 1984 *Physical Review B* 29 6443.
[15] M.S. Daw, M.I. Baskes 1983 *Physical Review Letters* 50 1285.
[16] M. Finnis, J. Sinclair 1984 *Philosophical Magazine A* 50 45.
[17] S.-G. Kim, M. Horstemeyer, M. Baskes, M. Rais-Rohani, S. Kim, B. Jelinek, J. Houze, A. Moitra, L. Liyanage 2009 *Journal of Engineering Materials and Technology* 131 041210.
[18] M.S. Daw, S.M. Foiles, M.I. Baskes 1993 *Materials Science Reports* 9 251.
[19] F. Ercolessi, J.B. Adams 1994 *EPL (Europhysics Letters)* 26 583.
[20] G.J. Ackland, M.I. Mendelev, D.J. Srolovitz, S. Han, A.V. Barashev 2004 *J. Phys.-Condes. Matter* 16 S2629.
[21] M.I. Mendelev, S. Han, D.J. Srolovitz, G.J. Ackland, D.Y. Sun, M. Asta 2003 *Philosophical Magazine* 83 3977.
[22] Y. Mishin, D. Farkas, M.J. Mehl, D.A. Papaconstantopoulos 1999 *Physical Review B* 59 3393.
[23] Y. Mishin, A.Y. Lozovoi 2006 *Acta Materialia* 54 5013.
[24] M.I. Mendelev, G.J. Ackland 2007 *Philosophical Magazine Letters* 87 349.
[25] Z. Wu, M. Francis, W. Curtin 2015 *Modelling and Simulation in Materials Science and Engineering* 23 015004.
[26] J. Hughes, M. Horstemeyer, R. Carino, N. Sukhija, W. Lawrimore II, S. Kim, M. Baskes 2015 *Jom* 67 148.
[27] M. Horstemeyer, J. Hughes, N. Sukhija, W. Lawrimore II, S. Kim, R. Carino, M. Baskes 2015 *Jom* 67 143.
[28] H.N.G. Wadley, X. Zhou, R.A. Johnson, M. Neurock 2001 *Prog. Mater. Sci.* 46 329.
[29] R.A. Johnson 1989 *Physical Review B* 39 12554.
[30] X. Zhou, H. Wadley, R.A. Johnson, D. Larson, N. Tabat, A. Cerezo, A. Petford-Long, G. Smith, P. Clifton, R. Martens 2001 *Acta Materialia* 49 4005.
[31] R. Ravelo, T.C. Germann, O. Guerrero, Q. An, B.L. Holian 2013 *Physical Review B* 88 134101(17 pp.).
[32] Y. Mishin, M.J. Mehl, D.A. Papaconstantopoulos, A.F. Voter, J.D. Kress 2001 *Physical Review B* 63 224106.
[33] J.H. Rose, J.R. Smith, F. Guinea, J. Ferrante 1984 *Physical Review B* 29 2963.

[34] M.J. Puska, R.M. Nieminen, M. Manninen 1981 *Physical Review B* 24 3037.
[35] M. Stott, E. Zaremba 1982 *Canadian Journal of Physics* 60 1145.
[36] A. Banerjea, J.R. Smith 1988 *Physical Review B* 37 6632.
[37] H.N.G. Wadley, X. Zhou, R.A. Johnson, M. Neurock 2001 *Prog. Mater. Sci.* 46 555.
[38] T. Wiss, R. Konings 2012 *Elsevier, Oxford* 465.
[39] K. Masuda, A. Sato 1981 *Philosophical Magazine A* 44 799.
[40] K. Masuda, K. Kobayashi, A. Sato, T. Mori 1981 *Philosophical Magazine B* 43 19.
[41] A.P. Sutton, M.W. Finnis, D.G. Pettifor, Y. Ohta 1988 *Journal of Physics C: Solid State Physics* 21 35.
[42] L. Wenhua, H. Wangyu, X. Shifang, D. Huiqiu, G. Fei 2010 *International Journal of Materials Research* 101 1361.
[43] M.D. Avinash, L. Bruce, L.I. Douglas, M.R. Arunachalam, A.Z. Mohammed, W.B. Donald 2012 *Modelling and Simulation in Materials Science and Engineering* 20 035007.
[44] A.M. Dongare, M. Neurock, L.V. Zhigilei 2009 *Physical Review B* 80 184106.
[45] C. Nan-xian, G. Xi-jin, Z. Wen-qing, Z. Feng-wu 1998 *Physical Review B* 57 14203.
[46] C. Nan-xian, C. Zhao-dou, W. Yu-chuan 1997 *Physical Review E* 55 R5.
[47] Y. Xiao-Jian, C. Nan-Xian, S. Jiang, H. Wangyu 2010 *Journal of Physics: Condensed Matter* 22 375503.
[48] K. Choudhary, T. Liang, A. Chernatynskiy, Z. Lu, A. Goyal, S.R. Phillpot, S.B. Sinnott 2015 *Journal of Physics: Condensed Matter* 27 015003.
[49] B. Jelinek, S. Groh, M. Horstemeyer, J. Houze, S. Kim, G. Wagner, A. Moitra, M. Baskes 2012 *Physical Review B* 85 245102.
[50] A.A. Lukyanov, S.B. Segletes 2012 *Applied Mechanics Reviews* 64 040802.
[51] A.A. Lukyanov 2008 *International Journal of Plasticity* 24 140.
[52] A. Dewaele, P. Loubeyre, M. Mezouar 2004 *Physical Review B* 70 094112.
[53] W.J. Nellis, J.A. Moriarty, A.C. Mitchell, M. Ross, R.G. Dandrea, N.W. Ashcroft, N.C. Holmes, G.R. Gathers 1988 *Physical Review Letters* 60 1414.
[54] P. Steve 1995 *Computational Materials Science* 4 361.
[55] P.J. Steinhardt, D.R. Nelson, M. Ronchetti 1983 *Physical Review B* 28 784.
[56] G. Woan, in: 2003, *CAMBRIDGE UNIVERSITY PRESS*.
[57] D.M. Eike, J.F. Brennecke, E.J. Maginn 2005 *The Journal of Chemical Physics* 122 014115.
[58] A. Stukowski 2012 *Modelling and Simulation in Materials Science and Engineering* 20 045021.
[59] M. de Koning, A. Antonelli, S. Yip 1999 *Physical Review Letters* 83 3973.
[60] M. de Koning, A. Antonelli 1996 *Physical Review E* 53 465.
[61] J. Mei, J.W. Davenport 1992 *Physical Review B* 46 21.
[62] D. Frenkel, B. Smit 2001 Understanding molecular simulation: from algorithms to applications Academic press.
[63] F.L. Tang, X.G. Cheng, W.J. Lu, W.Y. Yu 2010 *Physica B: Condensed Matter* 405 1248.
[64] R. Stedman, G. Nilsson 1966 *Physical Review* 145 492.
[65] A. Animalu, F. Bonsignori, V. Bortolani 1966 *Il Nuovo Cimento B Series 10* 44 159.

Table I. Potential parameters of aluminum, as well as Morse function parameters ($D_e$, $\alpha$, $r_e$), in present work. Note that the Morse parameters are the intermediate variables which can be determined by the potential parameters.

| Parameter | EAM-I | EAM-II | EAM-II' |
|---|---|---|---|
| $F_0$ | 2.699 978 11 | 2.699 997 35 | 2.699 995 08 |
| $n$ | 0.712 105 60 | 0.724 999 39 | 0.745 083 59 |
| $\theta$ | 2.009 549 60 | 1.886 118 85 | 1.693 915 00 |
| $kc$ | 0.635 610 00 | 0.692 050 00 | 0.362 560 00 |
| $\chi$ | 0.900 000 00 | 0.82 000 00 | 0.88 000 00 |
| $M_1^a$ | 0.838 465 64 | 0.687 681 14 | 0.717 746 11 |
| $M_2^a$ | -0.846 529 56 | -0.714 718 45 | -0.751 249 98 |
| $M_3^a$ | 0.061 826 87 | 0.070 601 33 | 0.026 767 01 |
| $M_4^a$ | 0.367 063 06 | 0.299 186 46 | 0.236 576 17 |
| | | | |
| $D_e$ | 0.107 910 28 | 0.165 040 66 | 0.131 979 48 |
| $\alpha$ | 1.113 131 35 | 0.874 169 42 | 1.074 277 11 |
| $r_e$ | 3.264 046 99 | 3.426 499 93 | 3.232 024 76 |

Table II. Quantities of Al predicted in present work in comparison with reference data. The quantities printed in bold have been contained in our reference database.

| Quantity | EAM-I | EAM-II | EAM-II' | Zhou et al.[a] | Mishin et al.[b] | Ref. |
|---|---|---|---|---|---|---|
| Lattice Properties | | | | | | |
| $a_0$ (Å) | 4.0496 | 4.0496 | 4.0496 | 4.08 | 4.05 | 4.0496[c] |
| $E_c$ (eV) | -3.36 | -3.36 | -3.36 | -3.58 | -3.36 | -3.36[b] |
| $E_{1v}$ (eV) | 0.66 | 0.66 | 0.68 | 0.71 | 0.68 | 0.66[b] |
| $B$ (GPa) | 79 | 79 | 79 | - | 79 | 79[b] |
| $C_{11}$ (GPa) | 114 | 114 | 114 | 127 | 114 | 114.0[d], 114.7[e] |
| $C_{12}$ (GPa) | 62.0 | 62.0 | 62.0 | 81.4 | 61.9 | 62.0[d], 61.24[f] |
| $C_{44}$ (GPa) | 31.7 | 31.6 | 31.7 | 36.4 | 31.6 | 32.0[d], 32.86[e] |
| $E$(BCC) (eV/atom) | -3.27 | -3.28 | -3.32 | -3.52 | -3.24 | -3.25[b] |
| $E$(HCP) (eV/atom) | -3.34 | -3.34 | -3.35 | -3.57 | -3.33 | -3.33[b] |
| Vibrational Properties | | | | | | |
| $v_L(X)$ (THz) | 9.52 | 9.50 | 9.48 | 8.51 | 9.31 | 9.69[b] |
| $v_T(X)$ (THz) | 5.87 | 5.86 | 5.90 | 5.65 | 5.98 | 5.80[b] |
| $v_L(K)$ (THz) | 7.77 | 7.76 | 7.78 | 6.79 | 7.30 | 7.59[b] |
| $v_{T1}(K)$ (THz) | 5.51 | 5.51 | 5.59 | 5.14 | 5.42 | 5.64[b] |
| $v_{T2}(K)$ (THz) | 8.63 | 8.62 | 8.63 | 7.72 | 8.28 | 8.65[b] |
| $v_L(L)$ (THz) | 9.70 | 9.71 | 9.70 | 8.86 | 9.64 | 9.69[b] |
| $v_T(L)$ (THz) | 4.08 | 4.12 | 4.24 | 4.16 | 4.30 | 4.19[b] |
| Stacking Faults | | | | | | |
| $\gamma_{SF}$ <112> (mJm$^{-2}$) | 119 | 114 | 30 | 70 | 146 | 166[f], 120-144[g,h] |
| $\gamma_{USF}$ <112> (mJm$^{-2}$) | 216 | 203 | 168 | 126 | 168 | - |
| Surface Energy | | | | | | |
| $\gamma_S$ (001) (Jm$^{-2}$) | 0.76 | 0.76 | 0.69 | 0.83 | 0.94 | 0.98[f] |
| $\gamma_S$ (110) (Jm$^{-2}$) | 0.97 | 0.99 | 0.92 | 1.16 | 1.01 | 0.98[f] |
| $\gamma_S$ (111) (Jm$^{-2}$) | 0.66 | 0.68 | 0.61 | 0.82 | 0.87 | 0.98[f] |

[a] Reference 30; [b] Reference 22; [c] Reference 51; [d] Reference 53;

[e] J.M. Winey, A. Kubota, Y.M. Gupta, Modelling and Simulation in Materials Science and Engineering, 18 (2010) 029801.

[f] L. E. Murr, Interfacial Phenomena in Metals and Alloys (Addison-Wesley, Reading, MA, 1975).

[g] Rautioaho R H 1982 Phys. Status Solidi b 112 83 and Westmacott K H and Peck R L 1971 Phil. Mag. 23 611

[h] K. H. Westmacott and R. L. Peck, Philos. Mag. 23, 611 ~1971.

Table III. Breakdown of contributions to free energy of liquid aluminum in equation (57), as well as the total free energy ($f_l$).

| Free energy | $\Delta f_1$ | $\Delta f_2$ | $f_{id}$ | $f_l$ |
|---|---|---|---|---|
| Value (eV) | -2.84328 | 0.18425 | -1.35747 | -4.01650 |

# Captions:

Fig. 1. Pairwise interactions of EAM potentials of aluminum

Fig. 2. Embedding energy as a function of $\rho/\rho_e$

Fig. 3. Atomic electron density function of EAM potentials of aluminum

Fig. 4. Predicted phonon dispersion curves at 0 K in comparison with the results from experiments: [a] Reference [64]; [b] Reference [65].

Fig. 5. Equation of states of aluminum predicted by different EAM potentials ([a] Ref. [30]; [b] Ref.[22]; [c] Ref. [51] ; [d] Ref. [52] )

Fig. 6. Bond parameters ($Q_6$) as a function of temperature, where the sudden drop indicates the occurrence of a melting transition.

Fig. 7. Snapshots of adaptive common neighbor analyses of simulated aluminum sample after relaxing for 80 ps at different temperatures covering its melting point. The potential of aluminum employed for the simulations is EAM-II. Similar results are observed by other potentials constructed in present work.

Fig. 8. W/N as a function of time for FCC aluminum and its liquid phase

Fig. 9. Comparison of $\langle U_1 - U_2 \rangle_\lambda$ calculated by reversible scaling (RS) technique and adiabatic switch (AS) approach at different λ, which suggests that either of the two methods could be employed for the evaluations of reversible work during adiabatic phase switch.

Fig. 10. (a) $\langle U - U_I \rangle_\lambda$ as a function of λ and (b) $P - P_I$ as a function of V for liquid phase of aluminum.

Fig. 11. Free energy as a function of temperature at zero extra pressure for FCC aluminum and its liquid phase.

Fig. 12. Equation of states changing with different single vacancy formation energy ($E_{1v}$) (a Ref. [51])

Fig. 13. Surface energy as a function of single vacancy formation energy ($E_{1v}$)

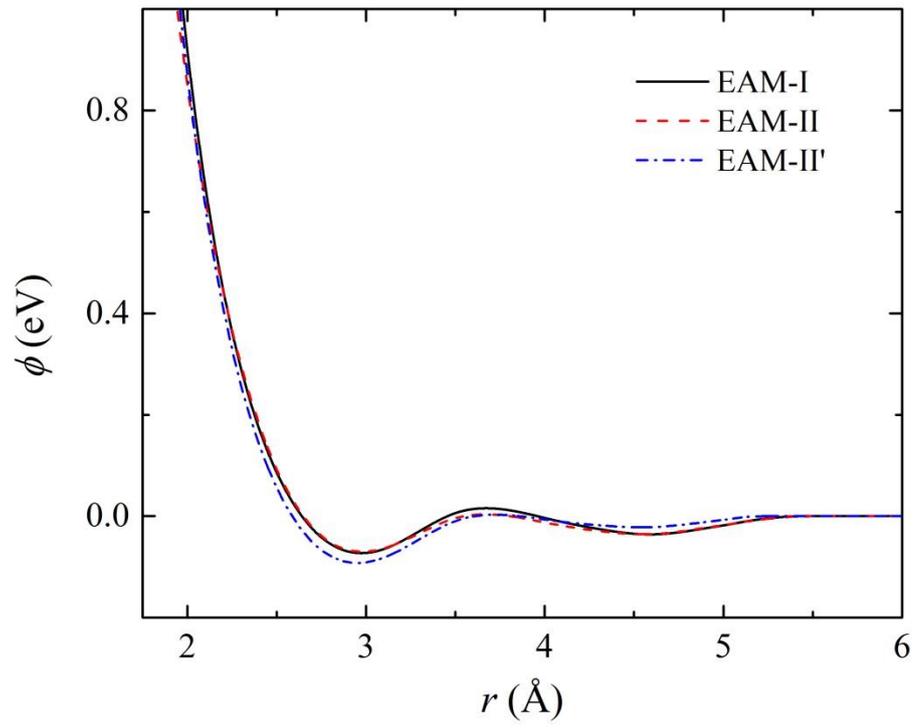

Fig. 1. Pairwise interactions of EAM potentials of aluminum

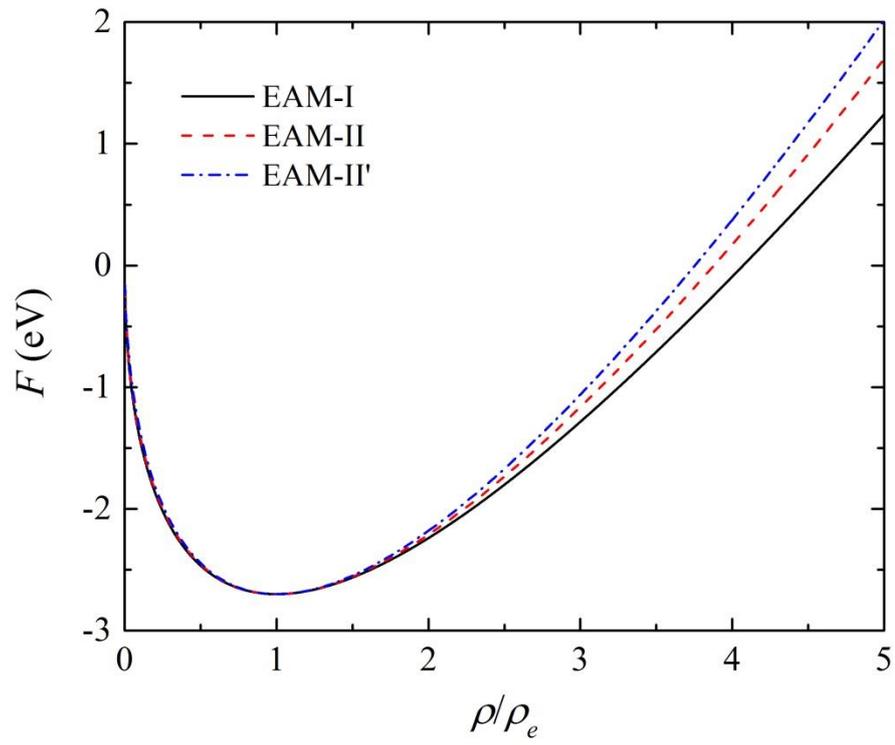

Fig. 2. Embedding energy as a function of $\rho/\rho_e$

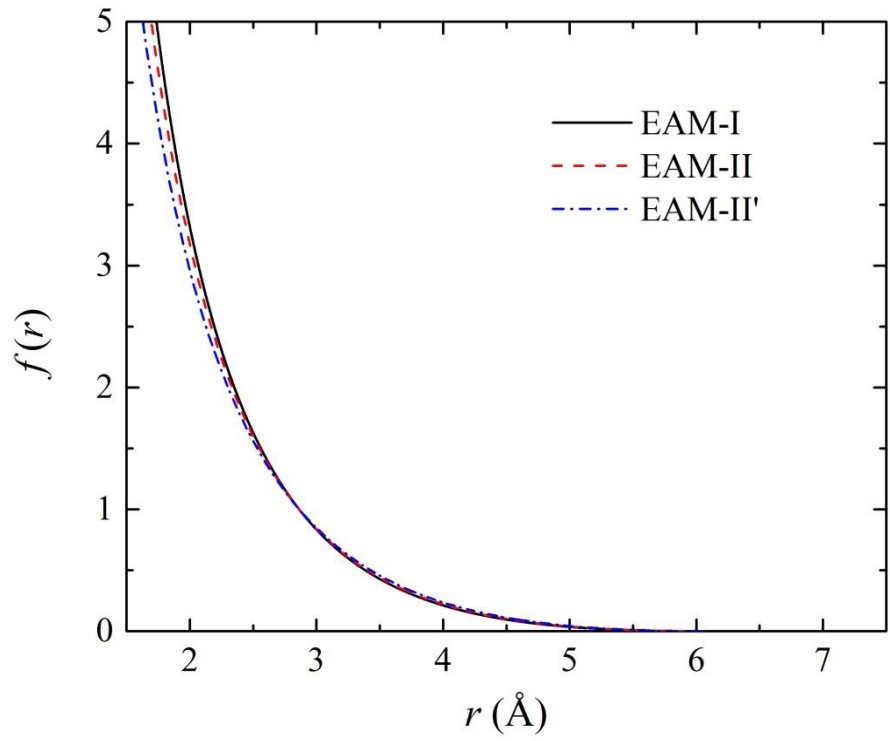

Fig. 3. Atomic electron density function of EAM potentials of aluminum

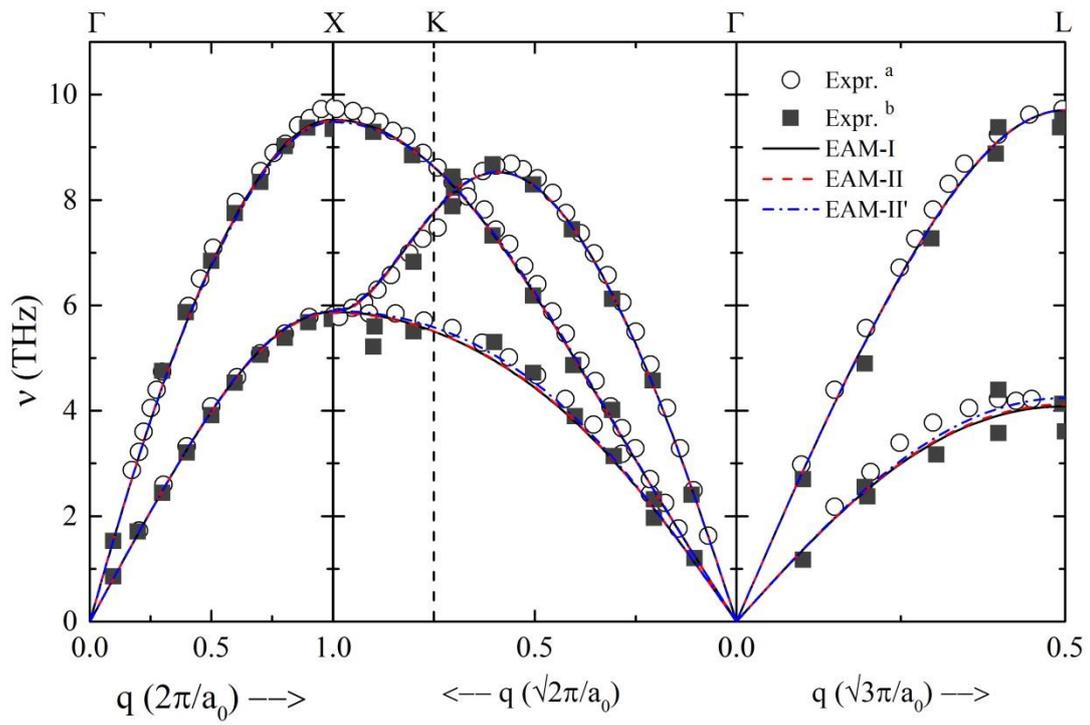

Fig. 4. Predicted phonon dispersion curves at 0 K in comparison with the results from experiments: [a] Reference [64]; [b] Reference [65].

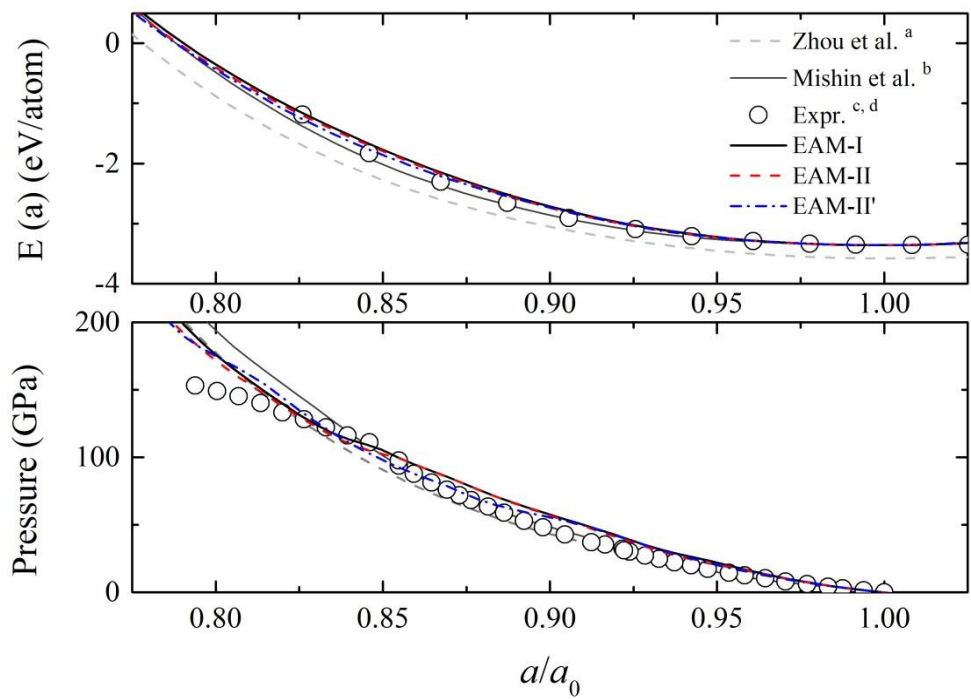

Fig. 5. Equation of states of aluminum predicted by different EAM potentials ([a] Ref. [30]; [b] Ref.[22]; [c] Ref. [52] ; [d] Ref. [53] )

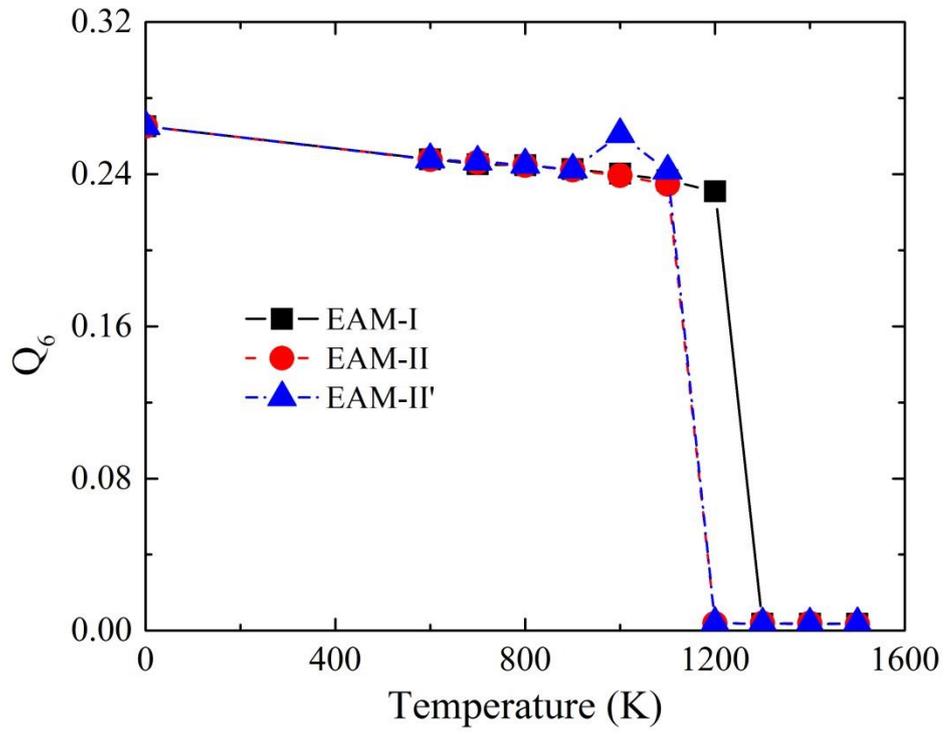

Fig. 6. Bond parameters ($Q_6$) as a function of temperature, where the sudden drop indicates the occurrence of a melting transition.

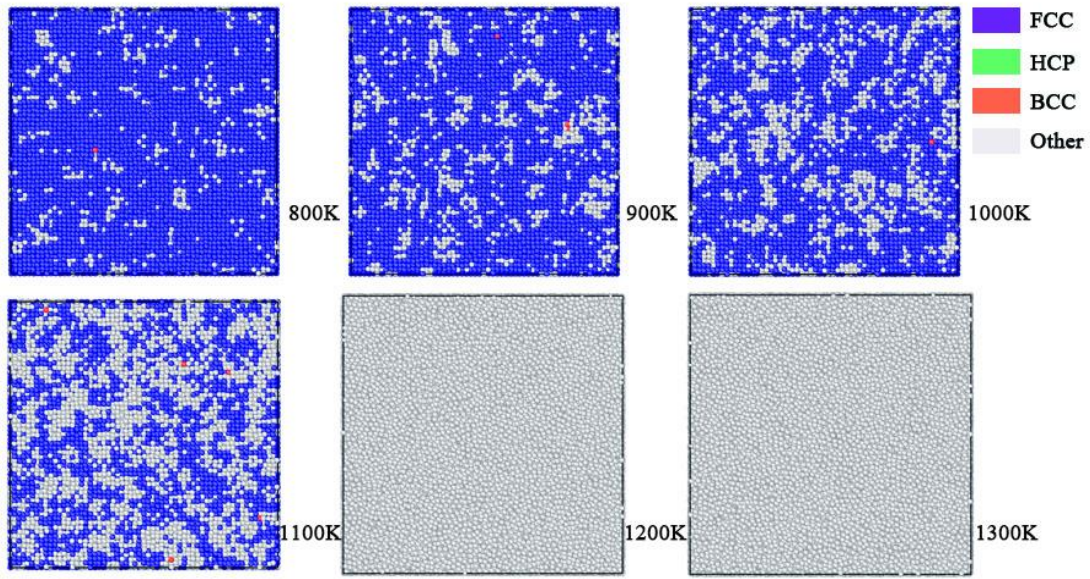

Fig. 7. Snapshots of adaptive common neighbor analyses of simulated aluminum sample after relaxing for 80 ps at different temperatures covering its melting point. The potential of aluminum employed for the simulations is EAM-II. Similar results are observed by other potentials constructed in present work.

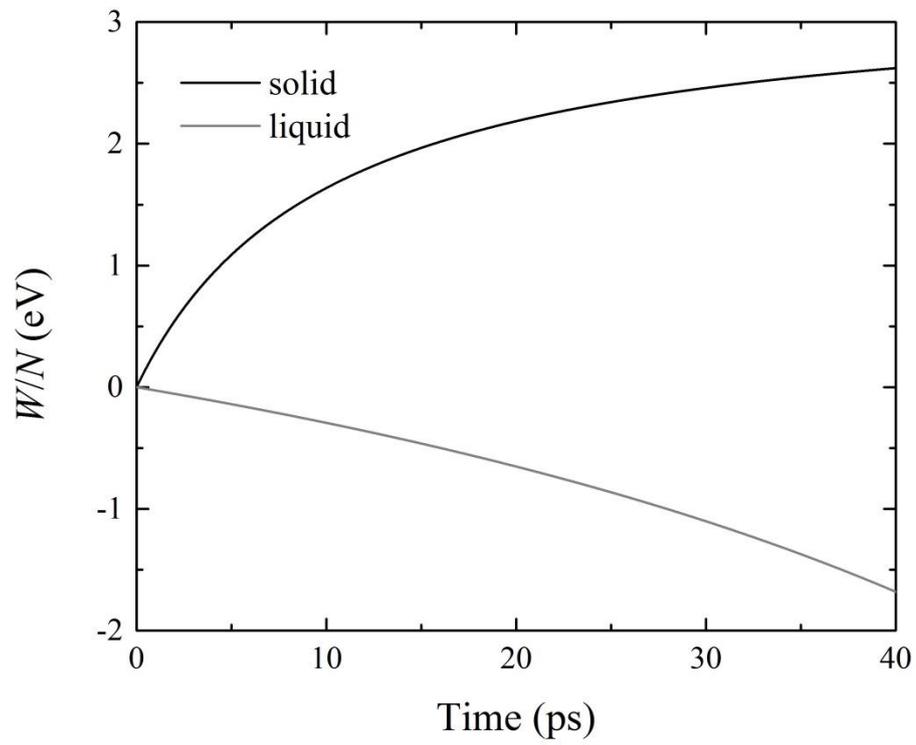

Fig. 8. *W*/*N* as a function of time for FCC aluminum and its liquid phase

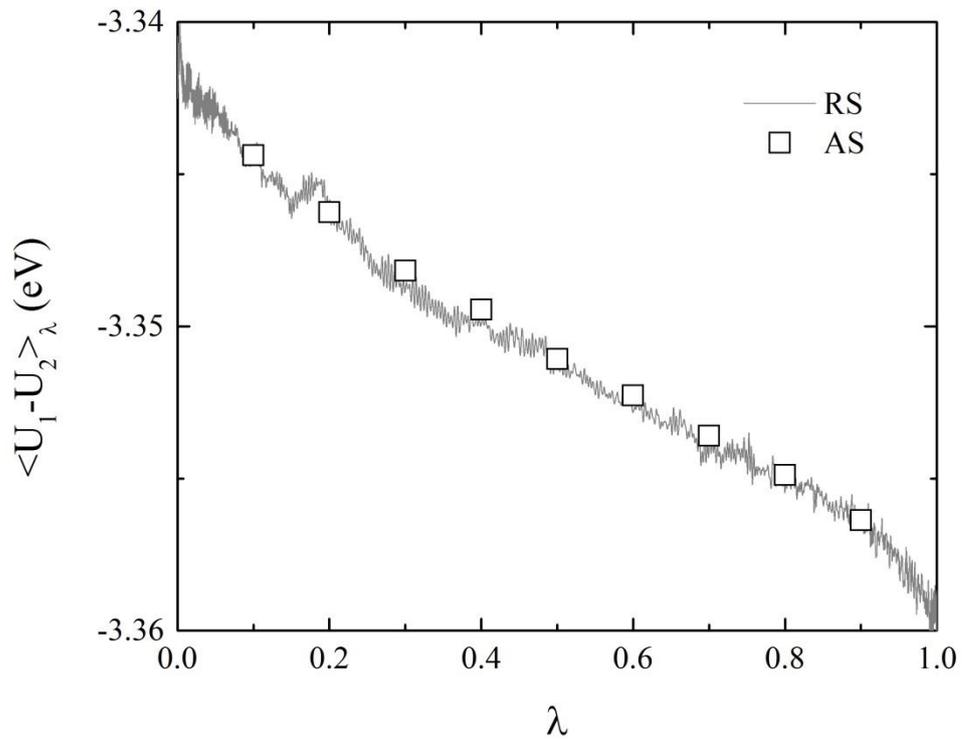

Fig. 9. Comparison of $\langle U_1 - U_2 \rangle_\lambda$ calculated by reversible scaling (RS) technique and adiabatic switch (AS) approach at different λ, which suggests that either of the two methods could be employed for the evaluations of reversible work during adiabatic phase switch.

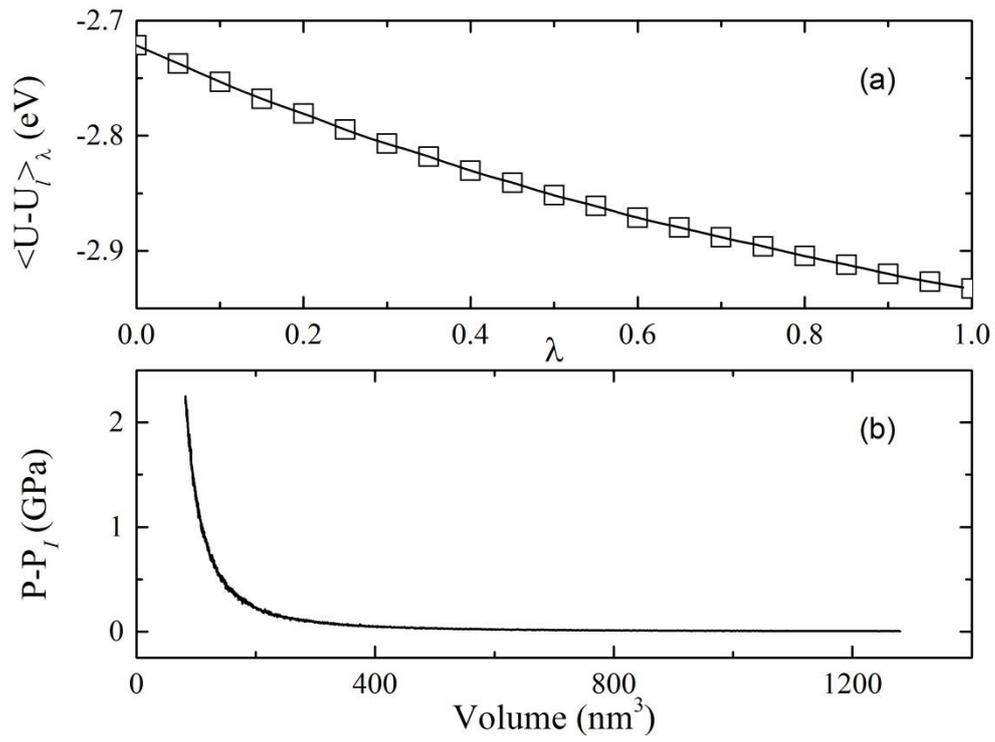

Fig. 10. (a) $\langle U - U_l \rangle_\lambda$ as a function of $\lambda$ and (b) $P - P_I$ as a function of $V$ for liquid phase of aluminum.

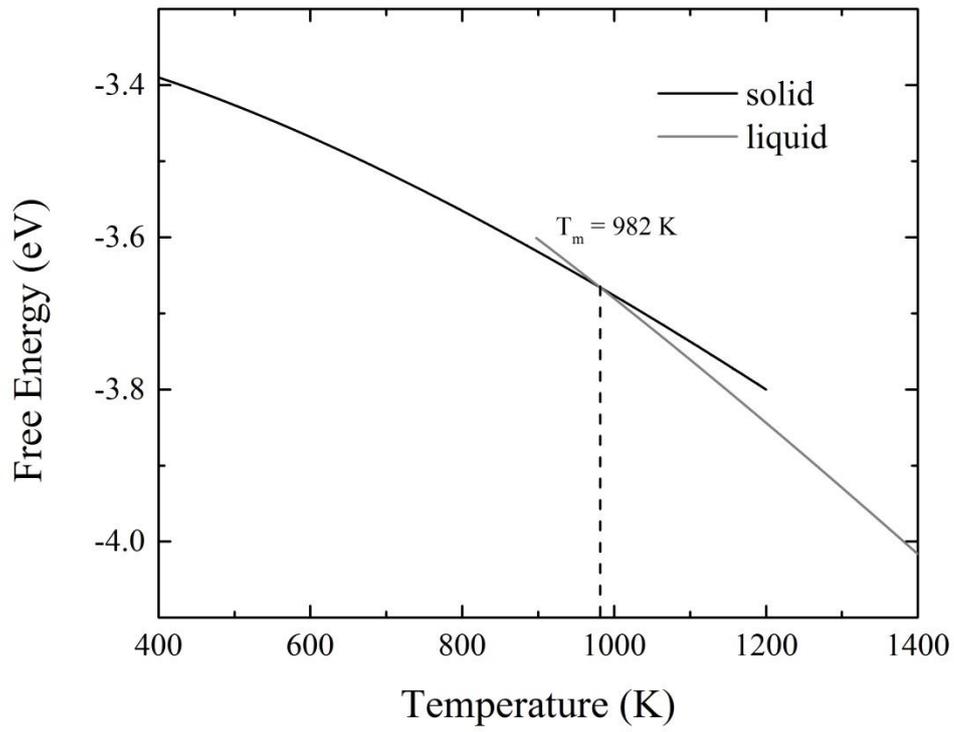

Fig. 11. Free energy as a function of temperature at zero extra pressure for FCC aluminum and its liquid phase.

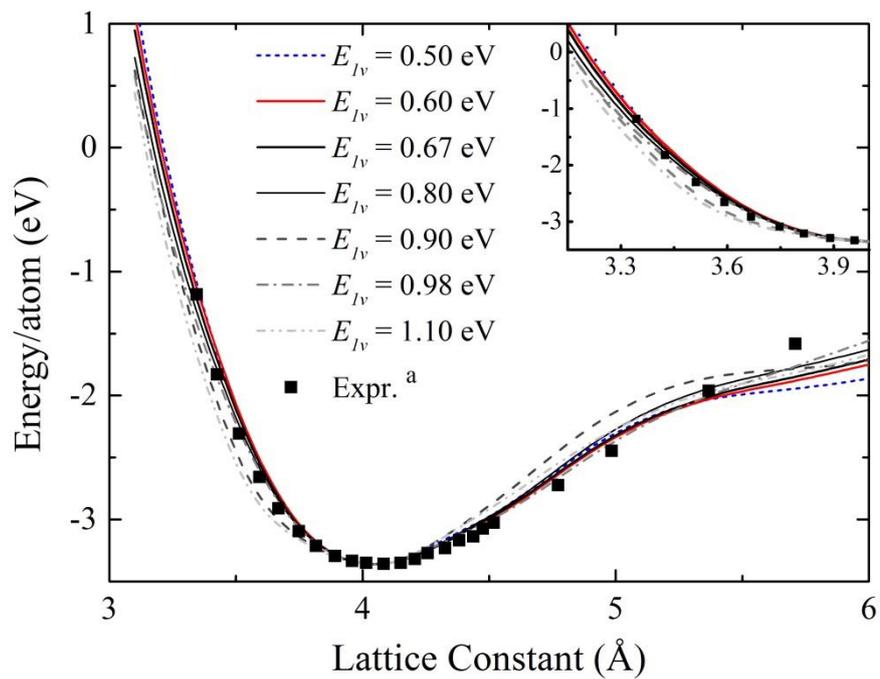

Fig. 12. Equation of states changing with different single vacancy formation energy ($E_{1v}$) ([a] Ref. [52])

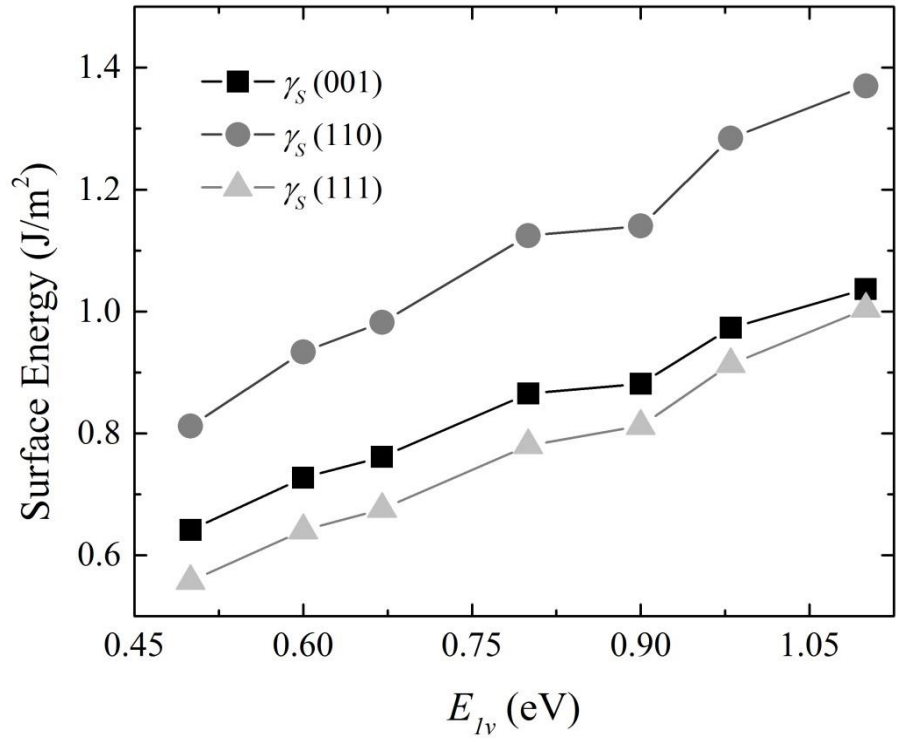

Fig. 13. Surface energy as a function of single vacancy formation energy ($E_{1v}$)